\documentclass{elsart}

\usepackage{natbib}

\usepackage{amssymb}
\usepackage{amsmath}
\usepackage{subfig}

\usepackage{graphics}
\usepackage{url}
\usepackage{color}

\definecolor{hcyan}{rgb}{0,0.66,0.92}
\definecolor{hblue}{rgb}{0.16,0.15,0.52}
\definecolor{hpink}{rgb}{0.87,0.06,0.5}
\definecolor{hred}{rgb}{0.87,0.06,0.15}
\definecolor{hgreen}{rgb}{0.,0.62,0.25}
\definecolor{hyellow}{rgb}{0.97,0.92,0.18}

\usepackage{algorithmic}
\usepackage{multirow}

\def\astrobj#1{#1}

\begin{document}
\newcommand{\nvidiaNS}{NVIDIA}
\newcommand{\nvidia}{NVIDIA }
\newcommand{\geforce}{GeForce }

\begin{frontmatter}



\thanks[label1]{Research undertaken as part of the Commonwealth Cosmology 
Initiative (CCI: www.thecci.org), an international collaboration 
supported by the Australian Research Council}
\title{Teraflop per second gravitational lensing ray-shooting using 
graphics processing units\thanksref{label1}}


\author{Alexander C. Thompson}, 
\author{Christopher J.\ Fluke\corauthref{cor1}},
\corauth[cor1]{Corresponding author. Tel.: +61 (0)3 9214 5828; fax: +61 (0)3 9214 8797.}
\ead{cfluke@swin.edu.au}
\author{David G.\ Barnes} 
and
\author{Benjamin R. Barsdell}

\address{Centre for Astrophysics and Supercomputing, 
Swinburne University of Technology, 
P.O. Box 218, Hawthorn, Victoria 3122, Australia}

\begin{abstract}
Gravitational lensing calculation using a direct inverse ray-shooting approach is 
a computationally expensive way to determine magnification maps, caustic 
patterns, and light-curves (e.g. as a function of source profile and size).  
However, as an easily parallelisable calculation, gravitational ray-shooting 
can be accelerated using programmable graphics processing units (GPUs).  
We present our implementation of inverse ray-shooting for the \nvidia G80 
generation of graphics processors using the \nvidia Compute Unified 
Device Architecture (CUDA) software development kit.  We also extend our 
code to multiple-GPU systems, including a 4-GPU \nvidia S1070 Tesla unit.  
We achieve sustained processing performance of 182 Gflop/s on a single GPU, 
and 1.28 Tflop/s using the Tesla unit.  We demonstrate that billion-lens 
microlensing simulations can be run on a single computer with a Tesla unit 
in timescales of order a day without the use of a hierarchical tree code.  
\end{abstract}


\end{frontmatter}

% main text
\section{Introduction}
\label{sct:intro}
Gravitational microlensing is the study of the deflection of light by matter 
in a regime where high magnification and multiple-imaging occurs, but
the individual micro-images are not resolvable.  This includes high 
magnification events due to lenses in the Galactic bulge and halo
(Alcock et al. 1993; Aubourg et al. 1993; Udalski et al. 1993),
and microlensing by compact objects within macro-lenses at cosmological
distances (Vanderriest et al. 1989; Irwin et al. 1989).  
While Galactic microlensing projects have focused on searches for dark 
matter and the detection of planets, cosmological microlensing 
has led to advances in the understanding of stellar mass functions, mean stellar 
masses, and the structure of quasars, including constraints 
on the physical size of the emission regions at different wavelengths. 
See Wambsganss (2006), Kochanek et al. (2007), Gould (2008), and Mao (2008) 
for recent reviews.

The standard signature of cosmological microlensing, especially when
applied to observations of active galactic nuclei,  is an 
uncorrelated change in brightness of a single macro-image within a 
multiply-imaged system (Schneider \& Weiss 1987). Intrinsic variation 
in source flux is seen as a correlated change in the brightness of all 
the images, separated by the (macro)lensing time-delay.  Such 
observations require accurate light curves to be obtained over long 
time periods, in many cases decades, as there is a wide variation in the 
time delay: $2-30$ hours for the quadruple-lensed \astrobj{Q2237+0305} 
(Vakulik et al. 2006) and 423 days for \astrobj{Q0957+561} 
(Hjorth et al. 2002) --  
see Saha et al. (2006) and Oguri (2007) for further examples.

Determination of the source size, source intensity profile, and physical 
properties of the microlenses (mass function, mean mass), requires a statistical 
comparison between observed light curves and microlensing models.  This is
achieved through the use of the gravitational lens equation:
\begin{equation}
{\mathbf y} = {\mathbf x} - \boldsymbol{\alpha} ({\mathbf x}),
\label{eqn:lens}
\end{equation}
which relates the two-dimensional locations of a source, ${\mathbf y}$, 
and an image, ${\mathbf x}$, with the deflection angle term, 
$\boldsymbol{\alpha}({\mathbf x})$, dependent on the arrangement of lenses.   
A common choice for microlensing is the many-Schwarzschild lens model:
\begin{equation}
{\boldsymbol \alpha} ({\mathbf x}) 
= \sum_{i=1}^{N_*} m_i \frac{({\mathbf x} - {\mathbf x}_i ) }
{\vert{\mathbf x} - {\mathbf x}_i \vert^2}
\label{eqn:slens}
\end{equation}
for $N_{*}$ lenses with masses, $m_i$, at positions ${\mathbf x}_i$.
The magnification, $\mu$, due to a gravitational lens system is   
\begin{equation}
\mu = 1/\det {\mathbf A}
\end{equation}
where ${\mathbf A} = \partial {\mathbf y}/\partial {\mathbf x}$ is the Jacobian
matrix of equation (\ref{eqn:lens}), 
which measures the areal distortion between the image and source
planes.
 
While an image position maps uniquely to a source location
(${\mathbf x} \rightarrow {\mathbf y}$ is a one-to-one mapping),
the converse is not true (${\mathbf y} \rightarrow {\mathbf x}$ is a one-to-many
mapping).  Except for a limited number of special cases [see
Schneider et al. (1992) for examples],  the lens equation is not
invertible.  In the cosmological microlensing case, where many
millions of individual stars may contribute to the observed magnification
of a macro-image,  it is more common to use a numerical technique to
solve for $\mu$ over a finite region of the source plane -- a magnification map --
rather than attempting to find all image locations from equation (\ref{eqn:lens}) 
for a given source position (e.g. Paczy\'{n}ski 1986).  

\begin{figure}
  \centering
  \scalebox{0.4}{\includegraphics{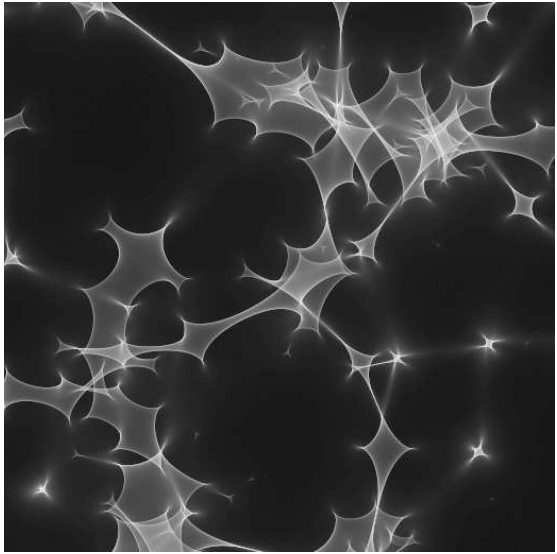}}
  \caption{A sample microlensing magnification map generated with a 
pair of \nvidia \geforce 8800GT graphics cards.  For model 
parameters: $N_{*} = 100$ lenses, $N_{\rm pix} = 1024^2$ pixels 
in the source plane, and $N_{\rm av} = 1000$ light rays per source pixel 
(on average), the processing time was $135$ s.  See section 
\ref{sct:implement} for details. }
\label{fig:map}
\end{figure}

Inverse ray-shooting provides the 
most straightforward means to obtain magnification maps for an arbitrary 
lens distribution [see Kayser et al. (1986) and Schneider 
\& Weiss (1986; 1987) for early versions of this technique].  
Inverse ray-shooting follows a large number (typically millions) of light rays
backwards from the observer, through the lens plane to the source plane, which
is represented as a pixellated grid.  The number of light rays falling in 
each pixel, $N_{ij}$, compared to the (average) number if there was no lensing,
$N_{\rm av}$, gives an estimate of the per-pixel magnification: 
\begin{equation}
\mu_{ij} = N_{ij}/N_{\rm av}.
\end{equation}
A typical magnification map is shown in Figure \ref{fig:map}, with the 
characteristic pattern of caustics clearly visible.  Caustics are regions of 
high magnification -- formally those points where $\det {\mathbf A} = 0$.  
The relative motion of the observer, lens plane and source imparts an
effective transverse velocity to the source, causing it to move across
the caustic network, and resulting in a time-varying change in source brightness.  
Accordingly, a sample light curve is generated by moving a source profile across 
a simulated caustic network, and converting the magnification at each point to 
a magnitude change.

Statistical investigations of cosmological microlensing require the generation
of many sample light curves, however, the creation of magnification maps 
poses a significant computational challenge.
The time to calculate a magnification map is directly proportional to the 
number of pixels in the source plane ($N_{\rm pix}$), the number of 
microlenses ($N_{*})$, and the number of floating point operations\footnote{We 
use the notation: flop = floating point operation; 
1 Gflop/s = 1 Gigaflop per second; and 1 Tflop/s = 1 Teraflop per second. } 
($N_{\rm flop}$) per deflection calculation.
As a Monte Carlo technique, the computation time is extended in direct 
proportion to $N_{\rm av}$, which sets the accuracy of calculated 
magnifications, and the number of repeat ($n$) map generations.  
Long compute times --- $O($days--months$)$ --- limit the scope to vary 
the input parameters, such as the initial stellar mass function, mean 
stellar mass, and source grid resolution. 
To keep computation times feasible for a direct implementation of
the inverse ray-shooting method, the product $\Phi = n \times N_{\rm flop} 
\times N_{\rm pix} \times N_{\rm *} \times N_{\rm av}$ 
historically has been constrained to $\lesssim O\left(10^{16}\right)$.

A number of approaches have been developed to overcome the processing time 
problem.  Wambsganss (1990; 1999) used a hierarchical tree code 
(Barnes \& Hut 1986), where lenses are treated differently depending on 
their distances from the light ray: lenses at a similar distance from a 
ray are grouped together and replaced with a single pseudo-lens of higher 
mass, effectively reducing the $N_*$ factor.  This introduces a slight error 
in the magnification map, which can be reduced by including higher order 
moments of the mass distribution.
A parallel version of the tree code, suitable for running 
billion-lens calculations on a parallel computing cluster -- a region
of parameter size previously unavailable to microlensing codes -- has been
implemented by Garsden \& Lewis (2009).

Mediavilla et al. (2006) used a lattice of polygonal cells to map areas of 
the image plane to source plane pixels, rather than using a regular
grid in the source plane.  This greatly reduces $N_{\rm av}$, resulting 
in a $\sim 100\times$ speed-up to reach a given accuracy compared to 
standard inverse ray-shooting, however, preparing an appropriate 
polygonal lattice introduces a significant computation overhead.

A limitation of Monte Carlo-style methods is that many more magnification 
values, $N_{\rm pix}$, are constructed than may actually be required 
(e.g. to form a single light curve).  A slightly larger (angular) 
size must be used for the image plane than for the source plane, as light 
rays at large impact parameters can be deflected into the source plane,
contributing flux that would otherwise be lost.  Consequently, more light rays 
must be generated than will actually fall within the source grid. Additionally, 
the finite source grid resolution means that true point source magnifications cannot be 
accurately calculated.  To avoid these issues, Lewis et al. (1993) 
and Witt (1993) independently developed approaches based on imaging 
an infinite line in the source plane,   which maps to a continuous, 
infinite line in the image plane, plus a number of closed loops - one 
for each microlens.   Wyithe \& Webster (1999) developed this technique 
further for extended sources.

In this work, we demonstrate that the redeployment of the direct
inverse ray-shooting algorithm on modern, programmable graphics
processing units (GPUs) can dramatically speed-up the calculation of
microlensing magnification maps, without the programming overheads of 
implementing a more complex code.  GPUs are macroscopic semiconductor arrays
designed to accelerate the rendering of three-dimensional geometry for 
display on two-dimensional computer screens.  Most modern computers 
contain a GPU, either on the system board or on a peripheral graphics card, 
which now regularly provide at least an order of magnitude greater raw 
computational power than the central processing unit (CPU).  
Rendering on-screen pixels is a highly parallel task, and this is
reflected in the GPU architecture.  Modern GPUs are primarily
composed of stream processors, which are individual arithmetic logic
units (ALUs) grouped in sets and controlled by an instruction
scheduler with associated shared memory.  
Consequently, algorithms that lend themselves to the ``stream
processing'' paradigm, where many individual data-streams all undergo
identical operations, can be moved to the GPU, resulting in
significantly shorter computation times.

On-going improvements in the performance of programmable graphics
hardware, combined with the notion of general purpose computation on
GPUs (GPGPU), is heralding a revolution in scientific computing
(Fournier \& Fussell 1988; Tomov et al. 2003; Venkatasubramanian 2003;
Owens et al. 2005).  
The two major graphics processor manufacturers, NVIDIA and AMD,
support GPGPU by providing software development kits for stream
computing.  NVIDIA introduced its Compute Unified Device
Architecture\footnote{\url{http://www.nvidia.com/cuda}} (CUDA) in 2006.  CUDA
is an extension (via compiler directives) of the standard C
programming language, designed to simplify the process of GPGPU
programming by abstracting the process of writing GPU code.
AMD's ``ATI Stream'' technology is based on an evolved form of
Brook.\footnote{\url{http://www-graphics.stanford.edu/projects/brookgpu/}} 
Brook was one of the first GPU languages that went
beyond the standard shader languages (particularly Cg\footnote{\url{http://www.nvidia.com}}
and  GLSL\footnote{\url{http://opengl.org}}) to
provide general purpose computing capabilities.  Both CUDA and ATI
Stream allow the programmer to define functions that are executed in
parallel on one or more GPUs attached to a system.

Astronomers are just beginning to consider the advantages that GPUs can 
offer for computation. Early work has focused on a few 
well-known algorithms, with gravitational N-body simulations receiving 
particular attention. Nyland et al. (2004) were the 
first to attempt to move the expensive $O(N^2)$ inter-body force calculations 
performed during an N-body simulation from the CPU to the GPU, work that was 
later followed by Portegies Zwart et al. (2007). 
Both groups found performance increases of $O(10)$ times, demonstrating the 
significant potential of GPUs, but their efforts were 
inhibited by the use of inflexible shader languages. The arrival of GPGPU 
languages improved the situation greatly, and measured performances of $O(100)$ 
times over CPU implementations led to a flurry of work in the area using 
BrookGPU (Elsen et al. 2007) and CUDA (Hamada \& Iitaka 2007; Belleman 
et al. 2007; Schive et al. 2007; Nyland et al. 2008; Moore et al. 2008). 
Other astronomy algorithms implemented on the GPU include radio-telescope 
signal correlation (Schaaf \& Overeem 2004; 
Wayth et al. 2007; Harris et al. 2008; Ord et al. 2009) and the solution of 
Kepler's equations (Ford 2008).  GPUs have also been used for real-time 
visualisation of large datasets (Szalay et al. 2008).

We now describe the implementation of GPU-based, direct inverse ray-shooting 
code for microlensing experiments. We demonstrate the substantial performance gains 
in executing our implementation on a system containing one or two 
\nvidia \geforce 8800 GT graphics cards, and on a system with an 
attached (external) \nvidia S1070 Tesla unit.  
The remainder of this paper is organised as follows.  
In section \ref{sct:implement} we describe 
both OpenMP (for multiple-core CPU systems) and our CUDA-based 
approach for single and multi-GPU systems.  We compare processing performance 
via timing tests in section \ref{sct:comparison}.  In section 
\ref{sct:discussion}, we consider advantages and limitations of 
GPGPU computing relevant to the ray-shooting case, and discuss 
applications of our approach, including computational steering. 

\section{Stream-processed ray-shooting}
\label{sct:implement}
The inverse ray-shooting algorithm is an ideal candidate for moving to a 
GPU as it is ``trivially parallelisable'': the deflection of a single light ray 
is independent of the deflection of all other light rays,  and the
deflection of a light ray due to one lens is independent of the deflection 
due to all other lenses.   The former case can be treated by splitting up the
total number of light rays, $N_{\rm ray}$, into $N_{\rm batch}$ batches
that are deflected in parallel.  The latter case can be treated by 
recasting the lens equation as:
\begin{equation}
{\mathbf x} - {\mathbf y} = \sum_{t=1}^{T} \sum_{i=1}^{N_t} m_i 
\frac{({\mathbf x} - {\mathbf x}_i ) }
{\vert{\mathbf x} - {\mathbf x}_i\vert^2}
\end{equation}
where $\sum_{t=1}^{T} N_t = N_*$ for $T$ parallel threads, and each 
processing step considers only a single light ray.  Our final solution uses
a combination of both parallel options.

The choice of parallel scheme necessitates a trade-off between
memory usage and processing speed. Aggregate GPU speed depends on the
following characteristics of the architecture:
the number of stream processors, which determines how many independent 
parallel tasks can be performed (per GPU stream clock cycle);
stream processor clock speed, which controls how many instructions
can be operated on in a given time period; 
memory bandwidth, which specifies how quickly memory can be accessed, 
and is typically faster on the GPU than the CPU;\footnote{e.g.
the \nvidia \geforce 8800 GT streams 57.6 GB/s, c.f. a typical
high-end CPU with dual-channel DDR2-800 with 12.8 GB/s of 
memory bandwidth} and GPU memory (classified as either device,
shared or register memory), which  
limits the amount of data that can be used for parallel processing.

\subsection{An OpenMP solution}
\label{sct:openMP}
As a step towards the GPU implementation, 
we consider OpenMP.\footnote{\url{http://openmp.org}}  
Processing threads are distributed 
amongst multiple CPUs on a single machine, with the run-time environment 
controlling thread allocation.  Compiler directives indicate code blocks to be 
processed in parallel, requiring minimal changes to the overall program 
structure.  In astronomy, OpenMP has been primarily used for gravitational
and hydrodynamical simulations (e.g. Thacker 1999; Semelin \& Combes 2005;
Merz, Pen \& Trac 2005; Thacker \& Couchman 2006; and Mudryk \& Murray 2009). 

The OpenMP direct ray-shooting algorithm works as follows:
\begin{enumerate}
\item $N_{*}$ lenses are generated on the CPU and stored in system memory.
\item $N_{\rm ray}$ light ray positions (${\mathbf x}$) are randomly generated 
in serial on the CPU and stored in system memory;  
\item Source coordinates are calculated in parallel, with light rays divided evenly 
between threads by OpenMP; 
\item Once complete, a single thread maps the ray locations onto the 
source pixel grid in order to obtain the magnification map.  
\end{enumerate}
This approach avoided potential problems such as multiple threads attempting 
to write to the same memory location when updating deflection angles, 
inaccurate or slow random number generation (the standard {\tt libc} random 
number generator {\tt rand()} is not thread safe), and enables a simpler extension 
to GPU programming with CUDA. We note that this algorithm assumes that all 
lens positions can be stored in CPU memory at one time, which limits
the range of $N_*$ that we can use for timing tests in section \ref{sct:timing}. 
\subsection{GPU ray-shooting with CUDA}
Our GPU ray-shooting algorithm operates in a similar manner to the 
OpenMP code, except that lens and light ray positions generated by the CPU and 
stored in the computer memory must be copied to and from the GPU's memory as 
part of each processing cycle.  For our specific implementation, we chose 
CUDA over ATI Stream technology for two main reasons: (i) it 
is available on more architectures (i.e. Windows, Linux and Mac OS/X, compared
to ATI Stream, which only supports Windows and Linux) 
and (ii) NVIDIA has been the first-to-market with desktop, GPGPU-specific, 
multi-Tflop/s products like the Tesla.
The C function calls of the OpenMP code are replaced with calls to 
the CUDA library.  

The GPU ray-shooting algorithm works as follows: 
\begin{enumerate}
\item $N_{*}$ lenses are generated on the CPU and loaded into GPU device memory;
\item $N_{\rm batch}$ light ray coordinates are randomly generated on the CPU and loaded into GPU device memory; 
\item GPU computation is initialised; computation is split into groups of 
128 threads (see below); 
\item Each thread group loads 128 lenses and calculates deflection on 128 rays, this is repeated 
until all lenses and rays are exhausted;
\item Once computation on the GPU is complete, results are copied
 back from the GPU device memory to system memory;
\item CPU maps the ray locations onto the source pixel grid in order to obtain the magnification map;
\item Steps 2-6 repeated until $N_{\rm av}$ rays per source pixel is reached. 
\end{enumerate}

Ray coordinates are generated and processed in batches, as this minimises memory usage
on the GPU while not reducing performance with a significantly large number of rays. 
For this work, $N_{\rm batch}$ was set to $2^{17}$, as larger sizes did not increase 
performance.
To maximise efficient use of the GPU device 
memory, source ray coordinates can overwrite the image 
ray coordinates, as they are still stored on the CPU memory. 
128 threads are run per group, which improves throughput, while not exhausting the supply 
of registers and shared memory. The GPUs on the Tesla unit have more registers,
allowing 256 threads per group. This number of 128 or 256 threads was chosen using
the \nvidia CUDA occupancy 
calculator\footnote{\url{http://developer.download.nvidia.com/compute/cuda/CUDA_Occupancy_calculator.xls}}
to determine best utilisation of the GPU, and this number was confirmed 
with testing.

\subsection{Multiple GPUs and the Tesla unit}
\label{sct:multigpu}
To make use of multiple GPUs or a Tesla device, multiple CPU threads are 
used. One CPU thread is associated with each GPU, handling memory transfer 
to and from the device,
and calls to device functions to perform computation.
For our implementation, CPU thread management was performed by OpenMP, as
it takes care of thread initialisation, destruction and synchronisation.
The multiple GPU algorithm works as follows:
\begin{enumerate}
\item $N_{*}$ lenses are generated on the CPU and loaded into all GPUs device memory;
\item $N_{\rm GPU} \times N_{\rm batch}$ light ray coordinates are randomly generated
in serial on the CPU by the master thread and stored in $N_{\rm GPU}$ arrays;
\item $N_{\rm GPU}$ threads copy ray coordinates to the GPUs and GPU 
computation is initialised; 
\item Computation is split into groups of 128 threads -- computation on 
each GPU is independent of the others; 
\item Each thread group loads 128 lenses and calculates deflection on 128 rays, this is repeated 
until all lenses and rays are exhausted;
\item Once computation on all GPUs is complete, results are copied
 back from the GPU device memory to system memory;
\item The master CPU thread maps the ray locations onto the
source pixel grid in order to obtain the magnification map;
\item Steps 2-6 repeated until $N_{\rm av}$ rays per source pixel is reached.
\end{enumerate}

While the best performance for our GPU algorithm was achieved 
when accessing only registers and shared memory, it 
is not possible to avoid either the transfer from CPU memory to GPU 
device memory, or from device to shared or register memory.  While this 
may seem like a processing bottleneck, the high latency of a single 
access of device memory is offset by the significant gain due to multiple 
streams being able to access device memory in parallel.  

\section{Comparing the approaches}
\label{sct:comparison}

\subsection{Timing Tests}
\label{sct:timing}
To evaluate the relative performance of OpenMP and CUDA inverse ray-shooting 
codes, a number of timing tests were performed.  We also implemented a 
single-CPU code: 
although the parameter space where this can be run is limited, it does provide
a check on the accuracy of the parallel codes. The same base hardware was used
throughout, comprising an Intel Q6600 Quad Core CPU (2.4 GHz), with 
4 GB of RAM and either two \nvidia GeForce 8800 GTs or an \nvidia S1070 
Tesla unit connected via dual PCIe x8, or higher, buses. In 
Table \ref{tab:devref}, 
we provide a summary of the hardware configurations. 

The 8800 GT comprises 112 stream processors, each running at 1.5 GHz, 
with a peak performance of 336 Gflop/s.\footnote{\nvidia quote a peak 
above 500 Gflop/s on this GPU, however, this assumes that MUL and MAD 
instructions can be dual-issued -- to our knowledge this is not possible 
with the 8800 GT.} This is much higher than a typical high-end 
CPU, such as the Intel Core 2 Quad 
Q6600, which is capable of 76.8 Gflop/s.  The 8800 GT has 512 MB of memory, 
which can store up to $6.7 \times 10^{7}$ two-dimensional lens positions as 
32-bit (i.e. single precision) floating point numbers.  The Tesla 
S1070\footnote{\url{http://www.nvidia.com/object/product_tesla_s1070_us.html}} 
comprises four GPUs, each with 240 stream processors, running at at 1.296 GHz. 
The Tesla's processing peak is 2.488 Tflop/s, and there is 4 GB available per
GPU.

\begin{table}
\caption{\label{tab:devref} Hardware configurations used in the comparison
of inverse ray-shooting codes. The memory column lists the total memory 
available for each processing option, which sets the problem size ($N_*$ and
$N_{\rm batch}$ light rays) that can be processed instantaneously. }
\begin{tabular}{llll}
{\bf Name} & {\bf Description} & {\bf Type} & {\bf Memory}\\
1CPU & Single thread on Intel Q6600 & CPU & 4 GB\\
4CPU & OpenMP with four threads on Intel Q6600 CPU & CPU & 4 GB\\
1GPU & Single \nvidia GeForce 8800 GT & GPU & 512 MB\\
2GPU & Two \nvidia GeForce 8800 GTs & GPU & 2$\times$512 MB\\
1TES & \nvidia S1070 Tesla with 4 GPU cores& GPU  & 4$\times$4 GB
\end{tabular}
\end{table}

\begin{figure}
  \centering
  \scalebox{0.25}{\includegraphics{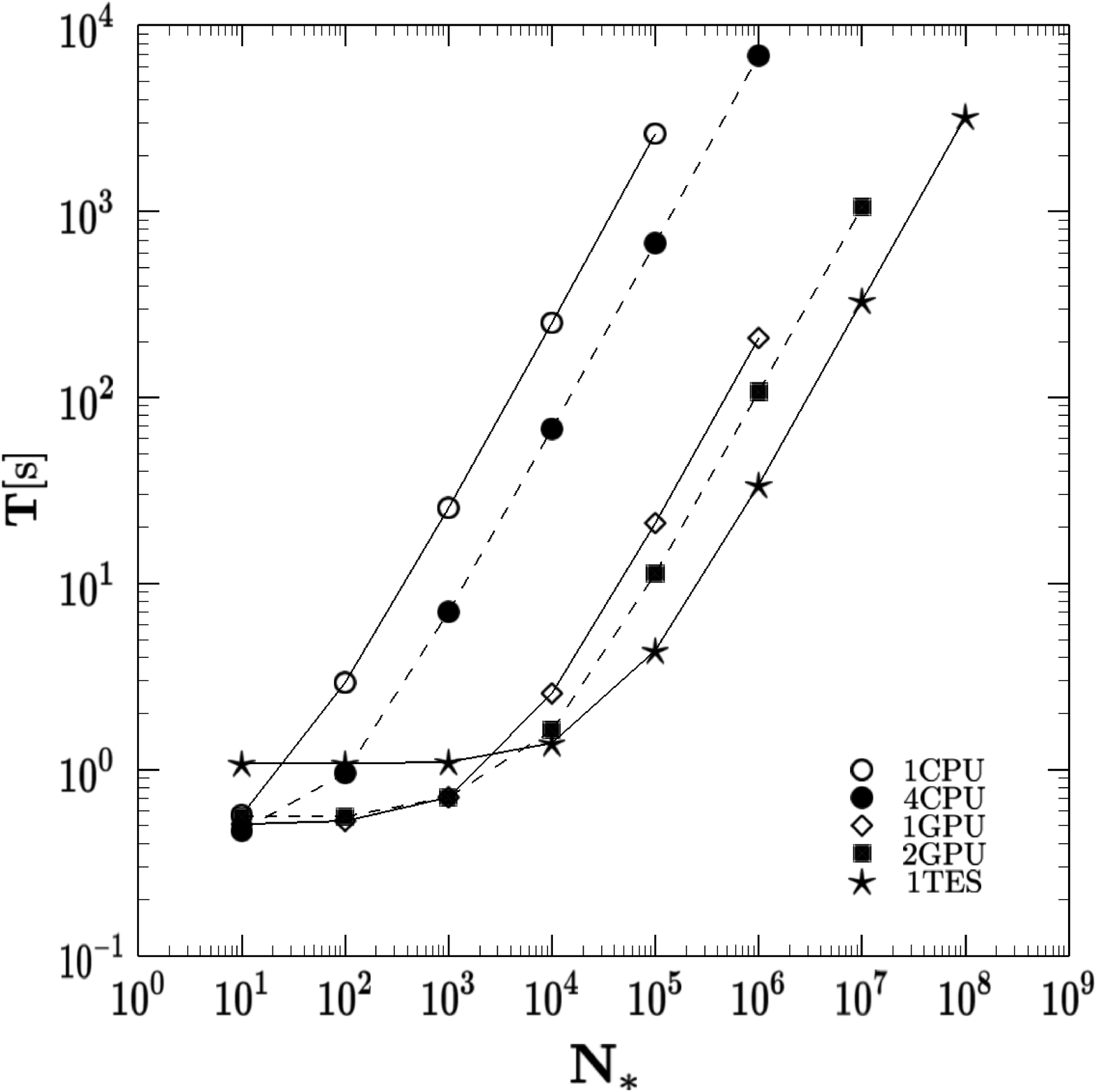}}
  \caption{Time ($T$ [s]) taken to generate magnification maps with 
each of the processing configurations. The number of lenses is varied over the range 
$10 \leq N_{*} \leq 10^{8}$, while $N_{\rm av} = 100$ and $N_{\rm pix} = 192^{2}$ 
are kept fixed for each run. Each plotted value is the median of three 
independent runs, variation between runs was below $0.5\%$.  Symbols 
are 1CPU (open circle), 4CPU (filled circle),
1GPU (open diamond), 2GPU (filled square) and 1TES (star). }
\label{fig:192loggraph}
\end{figure}

For each configuration, we run the relevant ray-shooting code with a fixed
set of parameters ($N_{*}, N_{\rm pix}, N_{\rm av}$) and random number
generator seed, which controls the locations of the lenses and the light rays.  
In order to maximise the memory available for lenses and light rays, we set
all lenses to have the same mass - see section \ref{sct:othermodels}.
We expect the accuracy of calculations with each code 
to be the same, as all use single precision floating point numbers. This was
confirmed by comparing the magnification maps for each run: we choose
the 1GPU map as our fiducial result, and determine the pixel-by-pixel residual 
for the other magnification maps. In each case, when the lens distributions
are the same and the light ray positions are the same, the residual is
zero for every pixel for each of the processing configurations.  This 
demonstrates
that our implementation of the algorithm across the GPU and Tesla hardware is
identical to the 1CPU version. An example map is shown in Figure \ref{fig:map} 
for $N_{*} = 100$, $N_{\rm pix} = 1024^2$ and $N_{\rm av} = 1000$. To 
obtain timing results, we run each simulation three times from a 
different random initial configuation, and quote the median processing time, 
determined by generating a time stamp at the start 
and end of processing. We find run-time variations of between $0.1-0.5\%$ across
all of the simulations.

Figure \ref{fig:192loggraph} show results of 
timing tests over a range of $N_{*}$. To keep 1CPU processing times 
manageable, we use a small source grid with $N_{\rm pix} = 192^2$ pixels and
$N_{\rm av} = 100$. 
Above $N_{*} = 100$, the 1CPU solution shows the expected linear 
scaling $\propto N_{*}$, as equation (\ref{eqn:slens}) is additive in 
the total number of lenses.  Due to the need to move data from CPU memory 
to GPU memory, the GPUs and Tesla have slightly higher overheads, which 
mean they have reduced performance for low $N_*$.  The linear scaling of
processing time with $N_*$ is not seen until $N_{*} \geq 10^{4}$ 
for 1GPU and $N_{*} \geq 10^{5}$ for 1TES.  

\begin{figure}[h]
  \centering
  \scalebox{0.25}{\includegraphics{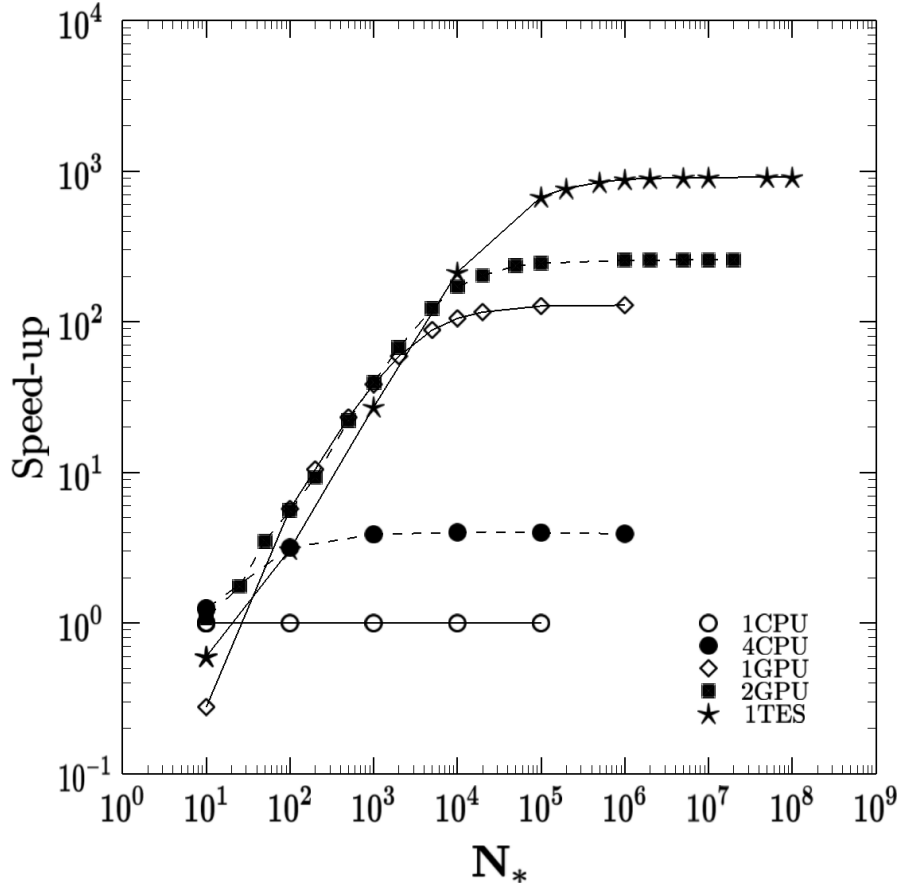}}
  \caption{Processing speed-up relative to 1CPU configuration as the number
of lenses is varied over the range $10 \leq N_{*} \leq 10^{8}$, 
with $N_{\rm av} = 100$ and $N_{\rm pix} = 192^{2}$. 
Each plotted value is the median of three independent runs, variation 
between runs was below $0.5\%$.  Symbols are 1CPU (open circle), 
4CPU (filled circle),
1GPU (open diamond), 2GPU (filled square) and 1TES (star). Peak performances
achieved were 1.4 Gflop/s (1CPU), 5.6 Gflop/s (4CPU), 182 Gflop/s (1GPU),
364 Gflop/s (2GPU) and 1287 Gflop/s (1TES).  Note that the 1CPU peak is 
assumed to apply beyond $N_* = 10^5$, and the maximum performance improvement 
for the GPU configurations occurs when $N_* \geq 10^5$. }
\label{fig:flops}
\end{figure}

Figure \ref{fig:flops} shows the relative speed-up in processing time compared
to the 1CPU configuration as the number of lenses is varied in the range 
$10 \leq N_{*} \leq 10^{8}$, with $N_{\rm pix} = 192^2$ and $N_{\rm av} = 100$. 
For each processing configuration, there is a problem-size where the performance
reaches a plateau level.  In calculating speed-ups for the GPU systems, we 
assume that the 1CPU peak applies beyond $N_* = 10^5$.  
At $N_{*} = 10^{4}$, 1CPU and 4CPU plateau at 1.4 Gflop/s and 5.6 Gflop/s 
respectively; while 1GPU, 2GPU and 1TES reach their peak performance 
of 182 Gflop/s, 364 Gflop/s and 1287 Gflop/s respectively for $N_{*} \geq 10^{5}$.
For low $N_*$, the extra overheads in transferring data to and from the GPU memory
impacts on the total processing time.  In order to achieve the greatest 
performance gains compared to the CPU implementation, GPUs need large problems 
to solve.

\begin{figure}[h]
  \centering
  \scalebox{0.25}{\includegraphics{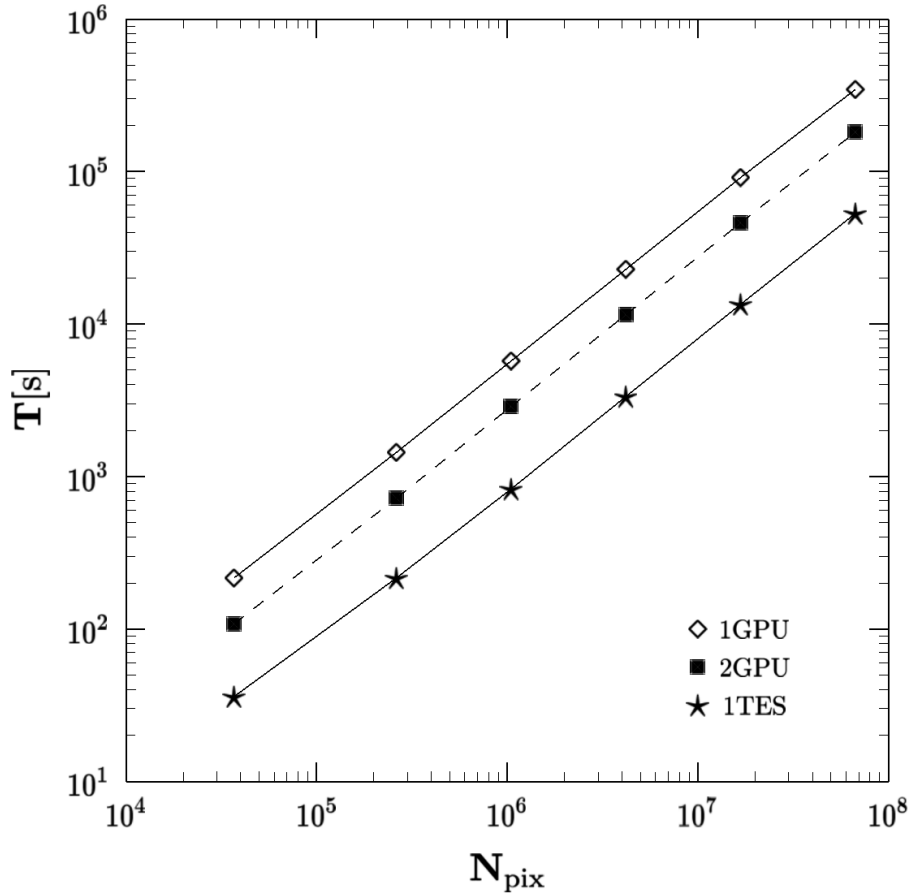}}
  \caption{Time ($T$ [s]) taken to generate magnification maps 
as the number of pixels in the source grid is varied over the range
$192^2 \leq N_{\rm pix} \leq 8192^{2}$.  The other lens parameters 
are kept fixed at $N_{\rm av} = 100$ and $N_{\rm *} = 10^{6}$.
Each plotted value is the median of three independent runs, variation 
between runs was below $0.5\%$.  Symbols are
1GPU (open diamond), 2GPU (filled square) and 1TES (star). }
\label{fig:grid}
\end{figure}

Table \ref{tab:gridSizes} and Figure \ref{fig:grid} show the processing times for calculating 
magnification maps over a range of grid sizes with $N_{*} = 10^{6}$.  
Numbers in [] brackets are estimated times for large maps which would 
be impractical to calculate with the 1CPU or 4CPU codes.  
The 2GPU code can calculate a $N_{\rm pix} = 4096^{2}$ pixel grid 
in 13 hours,  while 1TES can calculate a four-times larger grid 
in about the same time. While 1TES contains 4 GPUs, it can calculate 
more than 2x faster than 2GPU as each Tesla GPU gives approximately 
the same performance as 2GPU.
As with processing times for varying $N_{*}$, we find that processing
also scales linearly with $N_{\rm pix}$ as expected.

\begin{table}
\begin{center}
\caption{Time taken to generate magnification maps for each of the 
hardware configurations for varying $N_{\rm pix}$, with 
$N_{\rm av} = 100$, $N_{*} = 10^{6}$ fixed.  Numbers in [] brackets are 
estimated times for large maps which would be impractical to calculate.  Times
are in hours.}
\label{tab:gridSizes}
\begin{tabular}{cccccc}
\hline 
$N_{\rm pix}$ & 1CPU & 4CPU & 1GPU & 2GPU & 1TES \\
\hline
$192^2$  & 7.5     & 1.91  & 0.06   & 0.03  & 0.01   \\
$512^2$  & [55]    & [14] & 0.40   & 0.20  & 0.06  \\
$1024^2$ & [200]   & [55]  & 1.59   & 0.80  & 0.23  \\
$2048^2$ & [850]   & [200]  & 6.35   & 3.20  & 0.93  \\
$4096^2$ & [3400]  & [850]  & 25.4   & 12.8  & 3.74  \\
$8192^2$ & [13700]  & [3400]  & [100]   & 50.4  & 14.75 \\
\end{tabular}
\end{center}
\end{table}

\subsection{Other lens models}
\label{sct:othermodels}
A slight modification to the lens equation, which considers the effects of 
the macro-model shear, $\gamma$, and surface mass density, $\sigma$, is
\begin{equation}
{\mathbf y} = \left( 
\begin{array}{cc}
1 - \gamma & 0 \\
0 & 1 + \gamma 
\end{array}
\right) {\mathbf x} - \sigma_c {\mathbf x} 
- \sum_{i=1}^{N_*} m_i \frac{({\mathbf x} - {\mathbf x}_i ) }
{\vert{\mathbf x} - {\mathbf x}_i\vert^2}
\label{eqn:sigma}
\end{equation}
where $\sigma = \sigma_c + \sigma_*$ combines contributions from continuously
distributed matter, $\sigma_c$, and compact objects, $\sigma_*$.
Since shear and convergence effectively apply a scaling factor to the ray, which
can be performed as part of the accumulation stage, it neither significantly 
affects computation time nor requires increased memory usage.  
Ten years ago, Wambsganss (1999) stated that a direct inverse ray-shooting 
calculation for $N_{\rm pix} = 2500^2$, $N_{\rm av} = 500$ 
and $N_{*} = 10^6$, relevant for the astrophysically-motivated case 
of $\sigma_* \sim 1$ in equation (\ref{eqn:sigma}), was impractical 
as ``[e]ven with the fastest computers such a brute force calculation 
would take months or years!'' With an \nvidia S1070 Tesla unit, this 
brute force calculation can be achieved in $\sim 7$ hours. 

Of more importance
is the impact of allowing a variable lens mass, which does require more
memory usage: a third floating point number must be stored  for
the mass of each lens.  This means the maximum number of lenses that
a single GPU can store is decreased by 33\%.  While computation 
of variable lens mass does not take any additional floating point operations, 
it does require an additional memory look up per lens, increasing the total 
number of array look ups from two to three. Tests completed over a range 
of $N_{*}$ found a
run-time increase of approximately 17\% and 7\% for 1GPU and 1TES respectively.
The fact that a 50\% increase in memory did not lead to a 33\% drop in 
speed suggests that this computational problem is not memory bound.

\subsection{Billion-lens ray-shooting}
To compute $N_{*}$ greater than the available memory, it is necessary 
to partially calculate each light ray of a subset of lenses and sum 
the result after computation is complete across all lenses.  This can 
be approached in two ways, by loading and unloading lenses from GPU device 
memory, or by computing rays across multiple GPUs and transferring ray 
coordinates between GPUs.  Use of multiple GPUs is limited by the 
number of GPUs available and their device memory, while loading and 
unloading lenses is limited by system memory, storage space and overheads 
in transferring lens data to and from the GPU.  

We modify the algorithm of Section \ref{sct:multigpu} 
to stage data and pass light rays sequentially through
the series of 4 GPUs (numbered 1--4) in the Tesla unit.  Lens 
coordinates are pre-loaded 
onto each GPU, and the first batch of light rays is loaded from system memory  
to GPU-1. Once deflection calculations are completed on GPU-1, the 
partially-deflected light rays are transferred to GPU-2, and computation is
resumed.  Meanwhile, the next batch of light rays is loaded onto GPU-1.  
This staging process continues until all light ray deflections have been  
calculated, allowing the final accumulation process, and magnification 
map generation, to be completed on the CPU.  Staging ensures that there 
is always data ready to be processed and that the GPUs are kept utilised 
(except at the very start and end of the entire calculation, when light rays 
are first being passed to or from waiting GPUs). 

With this modification to our CUDA code, we ran a ray-shooting simulation
with $N_{*} = 10^{9}$, $N_{\rm pix} = 512^2$ and $N_{\rm av} = 42$. The
calculation was completed in 23.83 hours, with a sustained rate of 
1.28 Tflop/s. This indicates there is no significant overhead due to 
increased memory transfers and times when not all GPUs are being utilised.   
An additional test with $N_{*} = 2 \times 10^9$ was completed in 47.72 hours,
also at 1.28 Tflop/s, showing that processing time is still scaling 
linearly with $N_*$.

\section{Discussion}
\label{sct:discussion}

We have achieved 1.28 Tflop/s computation of gravitational lensing magnification 
maps by using a Tesla unit attached to a single desktop workstation.  For 
a single GPU we are reaching $54\%$ of the processing peak, 
and $52\%$ for the Tesla.  Other work, such as $N$-body simulations 
and graphic-card clusters for astrophysical simulations have achieved 
peak sustained performance of 223 Gflop/s (Levit \& Gazis 2006), 
236 Gflop/s (Belleman, B\'{e}dorf \& Portegies Zwart 2008) and 
256 Gflop/s (Hamada \& Iitaka 2007) with an \nvidia \geforce 8800 GTX card.  
The 8800 GTX has a peak of 346 Gflop/s, so these algorithms are using up to 74\% of 
available processing power.

It is desirable to quote the processing improvements that GPUs can 
provide over CPU implementations of astronomy algorithms, but care must be 
taken that the comparison is a fair and relevant one.  In this work, we 
have not considered the relative performance of GPU versus hierarchical 
tree-codes (Wambsganss 1990, 1999) or polygon mapping methods 
(Mediavilla 2006).  In order to isolate the performance enhancements of 
the GPU over a single CPU, we have deliberately used a simple ``brute 
force'' ray-shooting code, with no additional optimisation.   In future 
work, we will address comparisons between a cluster-based hierarchical
code (Garsden \& Lewis 2009) and the GPU approach for billion-lens 
configurations.

As with previous GPU-astronomy, the difference between the actual Gflop/s 
obtained and the quoted peak occurs due to the nature of astronomy algorithms:
in general they make use of more than just MAD instructions, which would be 
required to obtain maximum possible performance.
For the multi-Schwarzschild lens, we have $N_{\rm flop} = 10$ in our algorithm
analysis (see appendix \ref{app:flops}), with division 
counted as a single flop.  However, the calculation of division on an ALU 
takes $O(10)$ cycles, compared with a single cycle for MAD.
In addition to the computation,
the loop counter must be incremented and lens coordinates must be retrieved from
shared memory, adding a small overhead. Given these factors, it is not
surprising that we cannot reach the processing peak. 

We find that incremental improvements to the processing speed can be
made through judicious use of the device, shared and register memory.
To maximise performance, overheads from accessing device and
shared memory need to be minimised. One way to achieve this for device memory 
is via coalesced transfers, where there is a one-to-one mapping between 
threads and memory - each thread accesses its own unique device
memory location, up to the maximum number of threads.  This is then repeated 
until all the required device memory content has been transferred to shared or register
memory.  The GPU ray-shooting algorithm uses registers to store the ray coordinates
and shared memory to store the lenses. Since the ray coordinates are accessed
frequently it is advantageous to store them as registers as it allows maximum
throughput; lens coordinates are better stored in shared memory, so they can
be shared between threads, reducing the number of requests to device memory,
as they are only accessed once per thread.

Our comparison between GPU and CPU codes does not consider an obvious
optimisation that is reasonably straightforward on today's CPUs.
Modern CPUs typically have word widths of order 128 or 256 bits.  This
allows them (for example) to pass up to four single precision (32-bit)
floating point values through the processor on each clock cycle.  In
this example, single-instruction multiple data (SIMD) instructions
could be issued to the CPU's arithmetic logic unit to treat the
128-bit word as four single-precision floating-point numbers.  The
same floating point operation would be applied to each number, taking
the same number of clock cycles that an operation on just one number
would consume.  In the ideal case, this can quadruple the
floating-point speed of a {\em CPU-bound}\/ program.   However, the
{\em practical}\/ benefit of re-writing CPU code using SIMD extensions
is often less than a factor of four: simple mathematical operations
such as addition and multiplication are actually {\em memory
bandwidth-bound}\/ on contemporary architectures such as the Intel
Xeon E5345, so unless longer-cycle operations such as reciprocal and
reciprocal-square-root are in frequent use, the benefit of SIMD
operations is limited.

Most currently available GPUs have extremely limited ability to 
perform computation with double precision (64-bit) floating point numbers:  
the \nvidia G200 series and Tesla S1070 can typically only run double 
precision calculations at a fraction of the single precision rate (e.g. double
precision processing occurs at $\sim 300$ Gflop/s on the Tesla, 
compared to 2.488 Tflop/s for single precision). 
 Typical three-dimensional graphics processing does not require double 
precision, and as a result, this has not been a priority for hardware 
development.  Due to the stochastic nature of the inverse ray-shooting 
method (random lens positions and random ray positions), the additional
accuracy introduced by a move to double precision is not critical 
(however, this may become more of an issue when comparing with 
a hierarchical tree-code). As the market for GPGPU work grows, it is 
probable that (faster) double precision will become a standard 
feature.  In the meantime, careful scheduling of calculation can be 
used so that double precision is only used for those parts of the calculation 
that require it.

These discussions of algorithm comparisons notwithstanding, the processing 
speed-ups enabled with GPUs provides scope to consider improvements to the
way microlensing problem-space is addressed, and opportunities for 
real-time computation.  In section \ref{sct:intro}, we introduced $\Phi$ 
as a count of the number of calculations required for direct ray shooting. 
$\Phi$ is often kept manageable by foregoing any repeat map generations 
(i.e.\ $n = 1$), but this neglects the role of dynamic effects: 
static magnification maps do not account for the motion of individual 
stars, which can have a dramatic impact on the configuration of the 
caustic network over short timescales (Schramm et al. 1993), and may 
be a cause of rapid microlensing variability (Schechter et al. 2003; 
Kochanek et al. 2007).  An additional limitation of using a single 
magnification map is that all light curves are not statistically 
independent as source trajectories keep crossing the same set of 
caustics, however, it has not been feasible to produce a different 
magnification map for each sample light curve.  The order of magnitude 
improved performance of GPU-based systems over the CPU-only case
makes multiple-map generation (with direct calculations) a practical option 
-- even for very large $N_*$ scenarios.

Over a useful range of $N_*$ and $N_{\rm pix}$ values, we have demonstrated 
that the processing times for the 1GPU, 2GPU and 1TES solutions scale
linearly with these quantities.  Although we do not explicitly present 
results, we also find linear scaling in processing time for 
varying $N_{\rm av}$, in the regime where processors are fully utilised.
In order to examine the regions of cosmological microlensing parameter space 
that can be accessed with a GPU implementation of ray-shooting, we 
rewrite results from Table \ref{tab:gridSizes} as scaling relations:
\begin{eqnarray}
T_{\rm 1GPU} & = & 9.2 \times 10^4 \left(\frac{N_{\rm pix}}{4096^2}\right)
\left(\frac{N_{\rm *}}{10^6} \right) \left( \frac{N_{\rm av}}{100} \right)
\, \mbox{s} \\
T_{\rm 1TES} & = & 1.4 \times 10^4 \left(\frac{N_{\rm pix}}{4096^2}\right)
\left(\frac{N_{\rm *}}{10^6} \right) \left( \frac{N_{\rm av}}{100} \right)
\, \mbox{s}
\end{eqnarray}
If we consider a ``real-time'' calculation to be achievable in 60 seconds, then
the 1TES solution allows us to consider problems such as $N_{\rm pix} = 1024^2$,
$N_* = 10^5$ and $N_{\rm av} = 50$.  For an interactive, quick-look at
the effects of changing the lens mass distribution, or allowing time-evolving 
motion of lenses, computational steering with a direct ray-shooting calculation 
is feasible.  An alternative application is the ability to interactively zoom 
into a magnification map:  by initially starting with a low detail map, 
an area of interest can be selected and higher detail computed rapidly.

Finally, it is worth considering the financial cost of GPU-based
ray-shooting - although these figures are somewhat rubbery.  To 
achieve 1.2 Tflop/s with a cluster of 4CPU machines, with a peak 
of 5.6 Gflop/s, we would require $\sim 200$ four-core 
compute nodes.  Assuming perfect networking (at zero cost!), such a facility 
would cost $O(\sim 300$ 000$)$ USD, or around 250 000 USD per Tflop/s.  
The 2GPU solution costs $\sim 2000$ USD, with a peak of 364 Gflop/s.  Again,
with the assumption of perfect networking, we would need four machines, 
at a cost of $\sim 7000$ USD per Tflop/s.  This is comparable with the 
8000 USD per Tflop/s for 1TES (based on system price of 10 000 USD), with
no networking required.  It is obvious that on a financial
basis alone, the GPU-based solutions offer several orders of magnitude savings 
compared to CPU-clusters to achieve a particular processing goal.  What is not
yet known is how many other astronomy algorithms can benefit as significantly
from a GPU-implementation as the ray-shooting case -- this will be 
the subject of future work.

\section{Conclusions}
\label{sct:conclusion}

We have demonstrated the benefits of applying programmable graphics hardware
to a common problem in gravitational lensing: the creation of microlensing
magnification maps.  Multi-GPU ray-shooting can provide low-cost, Tflop/s 
performance without the programming complexities of implementing 
a hierarchical tree code, or the financial costs associated with 
using a distributed computing cluster.  
Significantly faster compute times expand the parameter space that can be 
explored in cosmological microlensing experiments, and open up new 
opportunities for ``real-time'' computational steering or interaction 
with a microlensing simulation.

\section*{Acknowledgements}
This research was supported under Australian Research Council's
Discovery Projects funding scheme (project number DP0665574). We thank
the anonymous referee for suggestions that improved the presentation of
results.


\appendix
\section{Counting flops}
\label{app:flops}
The many-Schwarzschild lens model, equation (\ref{eqn:slens}), 
requires $N_{\rm flop} = 10$ floating point operations per calculated deflection, 
counted as follows:

\begin{tabular}{lll}
$f_1$ & $= x_{1}-x_{i,1}$ & 1 flop\\ 
$f_2$ & $= x_{2}-x_{i,2}$ & 1 flop\\ 
$d$ & $= m_i/(f_1 \times f_1 + f_2 \times f_2)$ & 4 flops\\ 
$\alpha_1$ & $= \alpha_1 + (d \times f_1)$ & 2 flops\\ 
$\alpha_2$ & $= \alpha_2 + (d \times f_2)$ & 2 flops 
\end{tabular}

\end{document}